\documentclass[11pt,notitlepage,superscriptaddress, nofootinbib,longbibliography]{revtex4-1}
\usepackage{graphicx}
\usepackage{color}
\usepackage{amsfonts}
\usepackage{amsmath}

\newcommand{\comment}[1]{}

\begin{document}

\title{Asymptotic Floquet states of a periodically driven spin-boson system in the nonperturbative coupling regime}
\author{Luca Magazz\`u}
\affiliation{Institute of Physics, University of Augsburg, Universit\"atsstrasse 1, D-86135 Augsburg, Germany}
\author{Sergey Denisov}
\affiliation{Institute of Physics, University of Augsburg, Universit\"atsstrasse 1, D-86135 Augsburg, Germany}
\affiliation{Department of Applied Mathematics, Lobachevsky State University of Nizhny Novgorod, Nizhny Novgorod 603950, Russia}
\affiliation{Sumy State University, Rimsky-Korsakov Street 2, 40007 Sumy, Ukraine}
\author{Peter H\"anggi}
\affiliation{Institute of Physics, University of Augsburg, Universit\"atsstrasse 1, D-86135 Augsburg, Germany}
\affiliation{Nanosystems Initiative Munich, Schellingstra{\ss}e 4, D-80799 M\"unchen, Germany}

\date{\today}

\begin{abstract}
As a paradigmatic model of open quantum system, the spin-boson model is widely used in theoretical and experimental investigations.
Beyond the weak coupling limit, the  spin dynamics  can be described by a time-nonlocal generalized master equation with a memory kernel 
accounting for the dissipative effects induced by the bosonic environment. When the spin is additionally modulated by an external time-periodic electromagnetic field, the interplay between dissipation and  modulations yields a spectrum of nontrivial asymptotic states, especially in the regime of nonlinear response. 
Here we implement a method for the evaluation of Floquet dynamics in non-Markovian systems [L. Magazz\`u \emph{et al.}, Phys. Rev. A \textbf{96}, 042103 (2017)]  to find these strongly non-equilibrium states.
\end{abstract}


\maketitle

\section{Introduction}
\label{intro}

The spin-boson model~\cite{Leggett1987,*Leggett1987errata} has been -- and still remains -- the subject of extensive studies, as it constitutes an
archetype of open quantum system \cite{Cukier1989,Stockburger1996,Thorwart2009,Huelga2013,Han1991,Ustinov2018}. 
This model describes a single  spin (qubit)  coupled
to a dissipative environment consisting of an infinite number of bosonic modes. 
The spin-boson model is of relevance in a great variety of fields of physics and chemistry, with applications in electron tunneling~\cite{Cukier1989, Stockburger1996}, excitation transport in biological 
complexes~\cite{Thorwart2009, Huelga2013}, and superconducting qubit technologies~\cite{Han1991, Ustinov2018}, to name but a few. \\
\indent In the regime of weak coupling to the bosonic environment, the dynamics of the spin can be described by a master 
equation derived within the Born-Markov approximation~\cite{Petruccione2002} and improvements over this approximation, see for example Refs.~\cite{DiVincenzo2005,Schoeller2018}. 
Different techniques have been developed to deal with 
the more challenging coupling regimes which are beyond the perturbative limit. The current toolbox includes stochastic 
Schr\"odinger~\cite{Stockburger2002} and Liouville~\cite{Orth2013} equations, different variational 
approaches~\cite{Meyer1990,Wang2003,McCutcheon2011,deVega2017}, 
numerical renormalization group methods~\cite{Bulla2008}, sparse polynomial approach~\cite{Alvermann2009}, 
reaction coordinate ~\cite{Chin2010,Martinazzo2011,Iles-Smith2014} and combined thermofield/chain~\cite{deVega2015} mappings.\\
\indent Within the path integral approach, the influence of the environment on the evolution an open system (described with a reduced density matrix) 
is captured by the Feynman-Vernon influence functional~\cite{Weiss2012}. 
This formulation is suitable for numerically exact treatments such as Quantum Monte Carlo~\cite{Egger1992,Egger1994,Kast2013}, 
the quasi-adiabatic propagator path-integral method~\cite{Makarov1994,Makri1995,Thorwart2009}, and hierarchical equations 
of motion (HEOM)~\cite{Tanimura1989,Ishizaki2005, Tanimura2006,Moix2013}, 
but also provides an analytical tool
that can deal with the opposite limits of weak~\cite{Hartmann2000} and strong coupling~\cite{Leggett1987}, and even provides approximation schemes that bridge this two regimes~\cite{Nesi2007,Magazzu2015}.\\
\indent Different coupling regimes, corresponding different spin dynamics (coherent and incoherent), 
have been attained with superconducting qubits~\cite{Han1991,Chiorescu2003,Forn-Diaz2017, Ustinov2018}, currently
one of the  most popular platforms  for quantum computing and quantum simulations~\cite{You2011,Devoret2013, Wendin2017, Wilson2017}. 
In particular, a setup 
based on a flux qubit coupled to a transmission line has been proposed to  cover coupling strengths ranging from weak to ultra-strong~\cite{Forn-Diaz2017}.
The same setup has been recently used  to investigate the impact of a  monochromatic driving 
on the qubit dynamics in different dissipation and driving regimes~\cite{Magazzu2018}. 
Experimental results have been appropriately accounted for by the path integral approach 
within an approximation scheme which is nonperturbative with respect to the coupling strength. 
In this approach, the driven spin-boson dynamics is described with a generalized master equation (GME) 
for the spin, where a memory kernel constructed on a microscopical basis accounts for both the driving and dissipation~\cite{Grifoni1998}.\\
\indent While this GME for the spin dynamics cannot be evaluated within the standard Floquet formalism  (because of the time-nonlocal character of the dynamics), 
the evolution of the spin-boson system  in the full Hilbert space  is governed by a time-periodic Hamiltonian and, therefore, can be --
though only formally -- handled with the standard Floquet theory. 
The Floquet theory also applies to the 
reduced dynamics when the description can be given in terms of time-local evolution equations, 
typically in the weak coupling regime,  thus resulting in  the Bloch-Redfield or Lindblad master 
equations~\cite{Grifoni1998,Hausinger2010, Shirai2016,Hartmann2017,Bastidas2017}. 
In the limit of a high frequency driving, the Floquet approach can also be  implemented beyond the weak coupling limit ~\cite{Restrepo2016,*Restrepo2016errata},
by using the inverse frequency of modulations as a perturabtion parameter.\\
\indent In the present work we 
apply to the GME derived for the periodically driven spin-boson model the method presented in Ref.~\cite{Magazzu2017PRA}.
This method allows to find the asymptotic Floquet states of systems whose evolution is described by a master equations with memory kernels
and is based on a so-called 'embedding' procedure. The embedding is performed by coupling the  system 
to a set of non-physical variables, so that the resulting evolution equations governing the dynamics in the enlarged space of variables are  time-local. These equations have time-periodic coefficients and therefore are subjected to the standard Floquet theory. 
The projected solution in the physical  subspace has the form requested 
by a generalization of the Floquet theorem~\cite{Traversa2013} for systems with memory. 
We demonstrate that the  method allows to investigate the spin dynamics in different dissipation 
regimes under periodical modulations with no restrictions on their amplitude or  frequency. 

\section{Embedding and the generalized Floquet theory}\label{embedding}
In this section we outline a generalization of the  Floquet theorem~\cite{Floquet1883,Yakubovich1975} to systems with memory~\cite{Magazzu2017PRA}. 
Consider a periodically driven physical system  described by the ${\rm n}$-dimensional vector $\mathbf{x}(t)$ and governed by the time-nonlocal evolution equation
\begin{equation}\label{MKME}
\dot{\mathbf{x}}(t) =\int_{t_0}^{t}dt'\mathbf{K}(t,t')\mathbf{x}(t')+\mathbf{z}(t)
\end{equation}
with integrable memory kernel $\mathbf{K}(t,t')$. 
Assume that the  inhomogeneous term vanishes  asymptotically, i.e.,  $\lim_{t \rightarrow \infty} \mathbf{z}(t) =  \mathbf{0}$, 
and that the kernel matrix is bi-periodic, namely, $\mathbf{K}(t+\mathcal{T},t'+\mathcal{T}) =\mathbf{K}(t,t')$,  
where $\mathcal{T}$ is the period of the driving. 
Under these conditions, in the limit $t\rightarrow \infty$, the action of $\mathcal{L}\{\mathbf{x},t\}$, given by the right hand side of Eq.~(\ref{MKME}), 
commutes with that of the time-translation operator  $\mathcal{S}_{\mathcal{T}}\{\mathbf{x}(t)\}=\mathbf{x}(t+\mathcal{T})$. 
Then, according to the generalized Floquet theory~\cite{Traversa2013}, the solution  $\mathbf{x}(t)$ is formally expressed by
\begin{equation}\label{gFT}
\mathbf{x}(t) = \mathbf{S}(t,t_0)e^{(t-t_0)\mathbf{F}} \mathbf{v}(t_0)\;,
\end{equation}
with $\mathbf{S}(t,t_0)$ a $\mathcal{T}$-periodic ${\rm n} \times {\rm p}$ matrix, $\mathbf{F}$ a constant ${\rm p} \times {\rm p}$ matrix, 
and $\mathbf{v}(t_0)$ a ${\rm p}$-dimensional constant vector, where ${\rm p}\leq +\infty$.\\
\indent In Ref.~\cite{Magazzu2017PRA}, we presented a method to find the asymptotic solution of Eq.~(\ref{MKME}). The method is 
based on the enlargement of the system state space by coupling $\mathbf{x}$ to an auxiliary non-physical variable $\mathbf{u}$.
The resulting extended system is  described by the vector  $\mathbf{v}^{\mathsf{T}} =(x_1,\dots,x_{\rm n},u_1,u_2,\dots)$ which obeys a time-local equation. 
This equation is  then subjected to the standard Floquet treatment ~\cite{Yakubovich1975}. By projecting the  solution $\mathbf{v}(t)$ into the physical subspace, we 
find the solution for the original vector $\mathbf{x}(t)$.\\
\indent We assume that the ${\rm n}\times {\rm n}$ kernel matrix in Eq.~(\ref{MKME}) can be expressed as (or approximated by) the following sum of complex matrices
\begin{equation}\label{MK}
\mathbf{K}(t,t')=\sum_{j=1}^{{\rm k}}\Gamma_je^{-\gamma_j(t-t')}\mathbf{E}_j(t)\mathbf{F}_j(t')\;,
\end{equation}
with $\Gamma_j, \gamma_j\in\mathbb{C}$ and $\mathbf{E}_j(t) = \mathbf{E}_j(t+\mathcal{T})$,
$\mathbf{F}_j(t) = \mathbf{F}_j(t+\mathcal{T})$ complex  ${\rm n}\times {\rm n}$ matrices.
This form is flexible enough to reproduce -- at lest approximatively, as we do below -- a variety of memory kernels, including oscillatory ones~\cite{Beylkin2005}.\\
\indent With the kernel in the form of Eq.~(\ref{MK}), the time evolution of the physical variable  $\mathbf{x}$, Eq.~(\ref{MKME}),
can be obtained by solving the following  set of time-local equations~\cite{Kupferman2004} 
\begin{eqnarray}
\dot{\mathbf{x}}(t)&=&-\mathbf{H}(t)\mathbf{u}(t)\label{system1}\\
\dot{\mathbf{u}}(t)&=&-\mathbf{G}(t)\mathbf{x}(t)-\mathbf{A}\mathbf{u}(t)\;,\label{system2}
\end{eqnarray}
where the ${\rm n}$-dimensional vector $\mathbf{x}$ is coupled to the auxiliary variable $\mathbf{u}$. Here we have introduced the matrices
\begin{eqnarray}\label{matrices}
\mathbf{H}(t)&=&\begin{pmatrix}\Gamma_1\mathbf{E}_1(t)&\dots&\Gamma_{\rm k}\mathbf{E}_{\rm k}(t)\end{pmatrix}\;,\quad
\mathbf{G}(t)=\begin{pmatrix}
\mathbf{F}_1(t)\\
\vdots\\
\mathbf{F}_{\rm k}(t)
\end{pmatrix}\;,\nonumber\\
\text{and} \quad \mathbf{A}&=&\text{diag}(\gamma_1\mathbf{1}^{{\rm n}\times {\rm n}},\dots,\gamma_{\rm k}\mathbf{1}^{{\rm n}\times {\rm n}})\;.
\end{eqnarray}
These definitions entail that the vector of auxiliary variables has dimension  ${\rm n\cdot k}$. \\
\indent Without  loss of generality, we set $t_0=0$ from now on. To prove that Eqs.~(\ref{system1})-(\ref{system2}) 
are equivalent to Eq.~(\ref{MKME}), as far as the physical variable is concerned, we define $\mathbf{G}(t)\mathbf{x}(t)\equiv\mathbf{w}(t)$. 
Laplace transform of Eq.~(\ref{system2})  yields $\mathbf{u}(\lambda)=\left[ \lambda \mathbf{1}+\mathbf{A}\right]^{-1} \mathbf{u}(0)-\left[ \lambda \mathbf{1}+\mathbf{A}\right]^{-1} \mathbf{w}(\lambda)$.
Transforming back to the time domain, multiplying to the left by $-\mathbf{H}(t)$, and
using  Eq.~(\ref{system1}),  we recover Eq.~(\ref{MKME}) with 
\begin{equation}
\mathbf{K}(t,t')=\mathbf{H}(t) e^{-\mathbf{A}(t-t')}\mathbf{G}(t'),
\end{equation}
 [which is equivalent to Eq.~(\ref{MK})] provided that 
\begin{equation}
\mathbf{z}(t)=-\mathbf{H}(t) e^{-\mathbf{A}t}\mathbf{u}(0)\;.
\label{inhom}
\end{equation}
This requirement fixes the initial condition for the auxiliary vector $\mathbf{u}(t)$.\\
\indent Equations~(\ref{system1})-(\ref{system2}) can be cast in the compact form
\begin{eqnarray}\label{system-compact}
\dot{\mathbf{v}}(t)=\mathbf{M}(t)\mathbf{v}(t)\;,
\end{eqnarray}
where $\mathbf{v}^{\mathsf{T}}=(x_1,\dots,x_{\rm n},u_1,\dots,u_{{\rm nk}})$ is the ${\rm p}$-dimensional state 
vector associated to the enlarged system, with ${\rm p}={\rm n}+{\rm nk}$, and $\mathbf{M}(t)$ is a ${\rm p}\times{\rm p}$ matrix
with block structure
\begin{eqnarray}\label{M}
\mathbf{M}(t)=\begin{pmatrix}
\mathbf{0}&-\mathbf{H}(t)\\
-\mathbf{G}(t)&-\mathbf{A}\\
\end{pmatrix},
\end{eqnarray}
where $\mathbf{0}\in\mathbb{R}^{{\rm n}\times {\rm n}}$. Note that having $\mathcal{T}$-periodic $\mathbf{H}(t)$ and $\mathbf{G}(t)$ entails $\mathcal{T}$-periodicity of $\mathbf{M}(t)$. 
Then, according to the Floquet theorem, Eq.~(\ref{system-compact}) with initial condition $\mathbf{v}(t_0)$ has a formal solution whose projection in the physical subspace yields the form~(\ref{gFT})~\cite{Traversa2013, Magazzu2017PRA}. 
The asymptotic Floquet state of the original system at stroboscopic instances of time $t=s\mathcal{T},~s \in \mathbb{Z}$, is then given by the $\mathbf{x}$ component of the extended vector $\mathbf{v}^{\rm as}$, which is an invariant of the Floquet propagator $\mathbf{U}_{\mathcal{T}}=\hat{{\rm T}}\exp\left[\int_0^{\mathcal{T}}dt\;\mathbf{M}(t)\right]$, i.e., 
$\mathbf{U}_{\mathcal{T}}\mathbf{v}^{\rm as}(s\mathcal{T})=\mathbf{v}^{\rm as}(s\mathcal{T})$
~\cite{Yakubovich1975,Hartmann2017}; here  $\hat{{\rm T}}$ denotes the time-ordering operator.

\section{The periodically driven spin-boson model}
\label{model}
Now we apply the method described in the previous section to a periodically modulated spin-boson model. 
In this model~\cite{Leggett1987}, a two-state system, the spin (or qubit), interacts with 
a dissipative environment represented by a set  of independent bosonic modes of 
frequencies $\omega_i$ ('heat bath').  The strength of the 
coupling between the $i$-th mode and the two-state system is quantified by the frequency $\lambda_i$. According to the celebrated Caldeira-Leggett model~\cite{Caldeira1983}, to which we include
a time-dependent driving, the full Hamiltonian reads~\cite{Weiss2012}
\begin{eqnarray}\label{H}
H(t)&=&-\frac{\hbar}{2}\left[ \Delta \sigma_x+\varepsilon(t)\sigma_z\right]\nonumber\\
&&-\frac{\hbar}{2}\sigma_z\sum_i \lambda_i (a_i^{\dag}+a_i)+\sum_i\hbar\omega_i a_i^{\dag}a_i,
\end{eqnarray}
where  $\Delta$ is the bare transition amplitude per unit time between the eigenstates $|\pm1\rangle$ 
of the spin operator $\sigma_z$, and  $a_i$ ($a_i^{\dag}$) is the annihilation (creation) operator of the $i$-th bosonic mode.  

Following Ref.~\cite{Grifoni1998}, we consider a monochromatic driving of amplitude $\varepsilon_{\rm d}$ and frequency $\Omega$, with the time-dependent bias  of the form
\begin{equation}\label{drive}
\varepsilon(t)=\varepsilon_0+\varepsilon_{\rm d}\cos(\Omega t).\
\end{equation}
This setting corresponds to the setup used in recent experiments~\cite{Magazzu2018}.\\ 
\indent The bath and its coupling to the two-state system 
can be fully specified by the spin-boson spectral density function $G(\omega):=\sum_i \lambda_i^2\delta(\omega-\omega_i)$. 
For the bosonic heat bath, we assume the continuous Ohmic spectral density with a Drude cutoff~\cite{Weiss2012}
\begin{equation}\label{G}
 G(\omega)= 2\alpha \omega (1+\omega^2/\omega_{\rm c}^2)^{-1},
\end{equation}
 where the dimensionless parameter $\alpha$ quantifies 
 the overall system-bath coupling and $\omega_{\rm c}$ 
 is the cutoff frequency. In a physical system where the spin dynamics occurs through tunneling transitions between sites at distance $q_0$, the spin position operator is given by $\sigma_z q_0/2$.
Let $\rho(t)$ denote the spin density matrix. 
For a factorized system-bath initial condition at $t=0$, with the heat bath initially being in the canonical thermal state, 
the exact  dynamics of the population difference $\langle\sigma_z(t)\rangle=P_{+1}(t)-P_{-1}(t)$, where $P_{\pm 1}\equiv \langle \pm1|\rho(t)|\pm1\rangle$, 
is governed by the following GME:
\begin{equation}\label{GME}
\frac{d}{dt}\langle\sigma_z(t)\rangle=\int_{0}^tdt'\left[ K^{\rm a}(t,t')-K^{\rm s}(t,t')\langle\sigma_z(t')\rangle\right],
\end{equation}
where the symmetric (s) and antisymmetric (a) kernels -- with respect to $\varepsilon(t)$ --  have in general intricate path integral expressions~\cite{Grifoni1998,Weiss2012}.\\
\indent In the path integral picture, the Feynman-Vernon influence functional for the reduced system dynamics displays bath-induced, time-nonlocal interactions 
among the two-state transitions building up the paths. These interactions are mediated by the so-called pair interaction~\cite{Weiss2012}
\begin{eqnarray}\label{Q}
Q(t)&=&Q'(t)+{\rm i}Q''(t)\nonumber\\
&=&\int_0^\infty d\omega\;\frac{G(\omega)}{\omega^2}\left\{\coth\left(\frac{\beta\hbar\omega}{2}\right)\left[1-\cos(\omega t)\right]+{\rm i}\sin(\omega t)\right\}\;,
\end{eqnarray}
which is proportional to the second time integral of the bath correlation function.
In terms of Matsubara frequencies $\nu_k=2\pi k/\beta\hbar$ (with $k=1,2,\dots$), we get for the real part of the bath correlation function (see Ref.~\cite{Tanimura2006})
\begin{eqnarray}\label{ReQ}
\frac{d^2}{dt^2}Q'(t)&=&\int_0^\infty d\omega\; G(\omega)\coth\left(\frac{\beta\hbar\omega}{2}\right)\cos(\omega t)\nonumber\\
&=&c_0e^{-\omega_{\rm c}t}+\sum_{k=1}^\infty c_k e^{-\nu_k t}\;,
\end{eqnarray}
and thus
\begin{eqnarray}\label{Q2}
Q(t)=\sum_{k=0}^\infty \frac{c_k}{\nu_k}\left[t-\frac{1-e^{-\nu_k t}}{\nu_k}\right]+{\rm i}\pi \alpha\left(1-e^{-\omega_{\rm c}t}\right)\;,
\end{eqnarray}
where $\nu_0\equiv\omega_{\rm c}$ and where
\begin{eqnarray}\label{coefficients}
\frac{c_k}{\nu_k}&=&4\pi \alpha \frac{\omega_{\rm c}^2}{\nu_k^2-\omega_{\rm c}^2} \qquad(k\geq 1)\\
{\rm and}\qquad\frac{c_0}{\nu_0}&=&\pi\alpha\omega_{\rm c} \cot\left(\frac{\beta\hbar\omega_{\rm c}}{2}\right)\;.
\end{eqnarray}

For sufficiently high temperature and  cutoff frequency, $k_BT,\hbar\omega_{\rm c}\gtrsim \hbar \Delta$, 
the real part $Q'(t)$ becomes a linear function of time and and the imaginary part $Q''(t)$ is constant on a rather small time scale. 
In the path integral picture, this entails the decoupling of transitions which are distant in time. 
In this regime, the non-interacting blip approximation (NIBA), which  accounts for only local-in-time correlations~\cite{Leggett1987,Weiss2012}, very well 
describes  the dynamics of the population difference $\langle\sigma_z(t)\rangle$, as we show below. 
The NIBA is a nonperturbative approach with respect to the coupling $\alpha$: It is valid for the limit of strong coupling, when the stationary spin dynamics is fully incoherent.
Besides, it is accurate down to low temperatures, for every $\alpha$,
provided that the bias is zero, $\varepsilon(t)=0$.

Within the NIBA, the kernels $K^{\rm s/a}(t,t')$ in Eq.~(\ref{GME}) read~\cite{Grifoni1998}  
\begin{eqnarray}
\label{K}
K^{\rm s}(t,t')&=& h^{\rm s}(t-t')\cos[\zeta(t,t')]\nonumber\\
K^{\rm a}(t,t')&=& h^{\rm a}(t-t')\sin[\zeta(t,t')]\;,
\end{eqnarray}
with 
\begin{eqnarray}
\label{hsa}
h^{\rm s}(t-t')&:=&\Delta^2 e^{-Q'(t-t')}\cos[Q''(t-t')]\nonumber\\
h^{\rm a}(t-t')&:=&\Delta^2 e^{-Q'(t-t')}\sin[Q''(t-t')]\;.
\end{eqnarray}
The function $\zeta(t,t')=\int_{t'}^t dt''\varepsilon(t'')=\xi(t)-\xi(t')$, where
\begin{eqnarray}\label{xi}
\xi(t)=\varepsilon_0t+\frac{\varepsilon_{\rm d}}{\Omega}\sin(\Omega t)\;,
\end{eqnarray}
takes into account the modulations of the bias.
Note that, in the absence of a time-dependent driving, i.e.,  $\varepsilon_{\rm d}=0$, the GME  ~(\ref{GME})
with the NIBA kernels  given by Eq.~(\ref{K}) acquires a convolutive  character, namely the kernels depend exclusively on the difference $t-t'$.\\
\indent By using the definition $\langle\sigma_z(t)\rangle=P_{+1}(t)-P_{-1}(t)$ 
and the conservation of probability $P_{+1}(t)+P_{-1}(t)=1$,  we can cast Eq.~(\ref{GME}) into an equation desribing the evolution of the population vector $\mathbf{p}^{\mathsf{T}}=(P_{+1},P_{-1})$
\begin{equation}\label{GME2}
\dot{\mathbf{p}}(t)=\int_{0}^{t}dt'\mathbf{K}(t,t')\mathbf{p}(t')\;,
\end{equation}
which is of the form of Eq.~(\ref{MKME}) with dimension ${\rm n}=2$ and $\mathbf{z}(t)=0$. This form is suitable to be generalized to the case of a multi-site (tight-binding)~\cite{Grifoni1996, Grifoni1998} or multi-level~\cite{Thorwart2001,Magazzu2015, MagazzuJSTAT2016} system. These generalizations fall under the domain of applicability of the present treatment.  
The $2\times 2$ kernel matrix  $\mathbf{K}(t,t')$ in Eq.~(\ref{GME2}) has elements
\begin{eqnarray}\label{K3}
{\rm K}_{{+1-1 \above 0pt -1+1}}(t,t')&=&\frac{1}{2}\left[K^{\rm s}(t,t')\pm K^{\rm a}(t,t')\right]\nonumber\\
{\rm K}_{{+1+1 \above 0pt -1-1}}(t,t')&=&-{\rm K}_{{-1+1 \above 0pt +1-1}}(t,t')\;,
\end{eqnarray}
and can thus be written as
\begin{equation}\label{MK2}
\mathbf{K}(t,t')=\frac{1}{2}\sum_{j={\rm s,a}}h^{j}(t-t')\mathbf{E}_j(t)\mathbf{F}_j(t')\;.
\end{equation}
The effect of the driving is encapsulated in the time-dependent matrices $\mathbf{E}_j(t)$ and $\mathbf{F}_j(t)$ reading
\begin{eqnarray}\label{M}
\mathbf{E}_{\rm s}(t)&=&\begin{pmatrix}
\quad\cos[\xi(t)]&\quad\sin[\xi(t)]\\
-\cos[\xi(t)]&-\sin[\xi(t)]\\
\end{pmatrix}
\qquad \mathbf{E}_{\rm a}(t)=\begin{pmatrix}
\quad\sin[\xi(t)]&\quad\cos[\xi(t)]\\
-\sin[\xi(t)]&-\cos[\xi(t)]\\
\end{pmatrix}\nonumber\\
\mathbf{F}_{\rm s}(t)&=&\begin{pmatrix}
-\cos[\xi(t)]&\quad\cos[\xi(t)]\\
-\sin[\xi(t)]&\quad\sin[\xi(t)]\\
\end{pmatrix} 
\qquad\mathbf{F}_{\rm a}(t)=\begin{pmatrix}
\quad\cos[\xi(t)]&\quad\cos[\xi(t)]\\
-\sin[\xi(t)]&-\sin[\xi(t)]
\end{pmatrix}\;,
\end{eqnarray}
where the function $\xi(t)$ has been defined in Eq.~(\ref{xi}).\\
\indent In order to complete the embedding, we need to cast the kernel matrix in the form prescribed by Eq.~(\ref{MK}). To this aim, we approximate the two functions $h^{{\rm s/a}}(t)$ introduced in Eq.~(\ref{hsa}) as the following sums of oscillating exponentials 
\begin{eqnarray}\label{happrox}
h^{\rm s}(t)&\simeq& c_1 e^{-c_2t}\cos(c_3 t)+c_4 e^{-c_5t}\nonumber\\
&\equiv& 2\sum_{j=1}^{3}\Gamma_je^{-\gamma_j t} \nonumber\\
h^{\rm a}(t)&\simeq& e^{-d_1 t}\sin(d_2 t)+d_3\left(e^{-d_4 t}-e^{-d_5 t}\right)\nonumber\\
&\equiv& 2\sum_{j=4}^{7}\Gamma_je^{-\gamma_j t}\;. 
\end{eqnarray}
This yields the complex coefficients $\Gamma_j$ and $\gamma_j$ in Eq.~(\ref{MK}). Indeed, 
 by comparing with Eq.~(\ref{MK2}) we find
\begin{eqnarray}
\label{MKapprox}
\mathbf{K}(t,t')\simeq\sum_{j=1}^{7}\Gamma_je^{-\gamma_j(t-t')}\mathbf{E}_j(t)\mathbf{F}_j(t')\;,
\end{eqnarray}
with 
\begin{eqnarray}\label{coefficients}
&&\Gamma_1=\Gamma_2=c_1/4,\; \Gamma_3=c_4/2,\; 
\Gamma_{4}=-{\rm i}/4,\; \Gamma_{5}=+{\rm i}/4,\; \Gamma_{6}=+d_3/2,\; \Gamma_{7}=-d_3/2\nonumber\\
&&\gamma_1=c_2-{\rm i}c_3,\; \gamma_2=c_2+{\rm i}c_3, \gamma_3=c_5,\; 
\gamma_4=d_1-{\rm i}d_2,\; \gamma_5=d_1+{\rm i}d_2,\; \gamma_{6}=d_4,\; \gamma_{7}=d_5\nonumber\\
\end{eqnarray}
and 
\begin{eqnarray}\label{matricesEF}
&& \mathbf{E}_1(t)=\mathbf{E}_2(t)=\mathbf{E}_3(t)=\mathbf{E}_s(t),\quad\mathbf{E}_4(t)=\mathbf{E}_5(t)=\mathbf{E}_6(t)=\mathbf{E}_7(t)=\mathbf{E}_a(t)\nonumber\\
&&\mathbf{F}_1(t)=\mathbf{F}_2(t)=\mathbf{F}_3(t)=\mathbf{F}_s(t),\quad \mathbf{F}_4(t)=\mathbf{F}_5(t)=\mathbf{F}_6(t)=\mathbf{F}_7(t)=\mathbf{F}_a(t)\;.
\end{eqnarray}
Once that the physical parameters of the problem are fixed, the coefficients $c_j,d_j\in \mathbb{R}$ can be obtained upon a fitting procedure, see Appendix~\ref{fit}, by using the approximated expressions~(\ref{happrox}) as fitting functions. The coefficients, in turn, produce the complex coefficients $\gamma_j$ and $\Gamma_j$ through Eq.~(\ref{coefficients}). 
Applying to the present case the general rule, the dimension of the embedded system is $\rm n+{\rm nk}=16$ with $\rm n=2$, the dimension of the physical variable $\mathbf{p}$, and ${\rm nk}=14$ the dimension of the auxiliary variable $\mathbf{u}$, being $\rm k=7$ (see Eq.~(\ref{MKapprox})).
The extended system is thus described by $\mathbf{v}^{\mathsf{T}}=(P_{+1},P_{-1},u_1,\dots,u_{14})$. Its asymptotic state is invariant under the action of the Floquet propagator $\mathbf{U}_{\mathcal{T}}$. The latter is in practice constructed as follows. Its columns are the one-period  propagated elements of the canonical basis of $\mathbb{R}^{16}$ $(1,0,\dots,0),\;(0,1,\dots,0),\dots,(0,0,\dots,1)$. The propagation is performed via the time-local matrix equation~(\ref{system-compact}) by using a $4$-th order Runge-Kutta scheme with the rather small time step $\delta=\mathcal{T}/10^{4}$, where $\mathcal{T}=2\pi/\Omega$.
The invariant vector $\mathbf{v}^{\rm as}$ is the eigenvector of 
$\mathbf{U}_{\mathcal{T}}$ corresponding to eigenvalue $1$. The asymptotic Floquet state on a time span of one period is then obtained by propagating for one period the extended system, with $\mathbf{v}^{\rm as}$ as initial condition.\\
\indent In concluding the present section we note that if the driving period $\Omega$ is a multiple of the static bias $\varepsilon_0$, then the extended system has the period $\mathcal{T}=2\pi/\Omega$ of the driving. Moreover, if the ratio $\Omega/\varepsilon_0$ is a rational number then the periodicity of the extended system is  
$\mathcal{T}=m 2\pi/\varepsilon_0$, where $m$ is the minimum integer such that the ratio 
$m\Omega/\varepsilon_0$, is also an integer. On the other hand, if 
the frequencies $\varepsilon_0$ and $\Omega$ are not commensurate, then the embedding procedure 
still yields a time-local system but with non-periodic coefficients; such systems do not allow for the Floquet treatment. 
However, these considerations hold for the extended system while the physical degrees of freedom have, asymptotically, the periodicity of the driving, 
independently of the values of static bias and driving frequency.

\section{Results}

\begin{figure}[ht!]
\begin{center}
\includegraphics[width=.95\linewidth,angle=0]{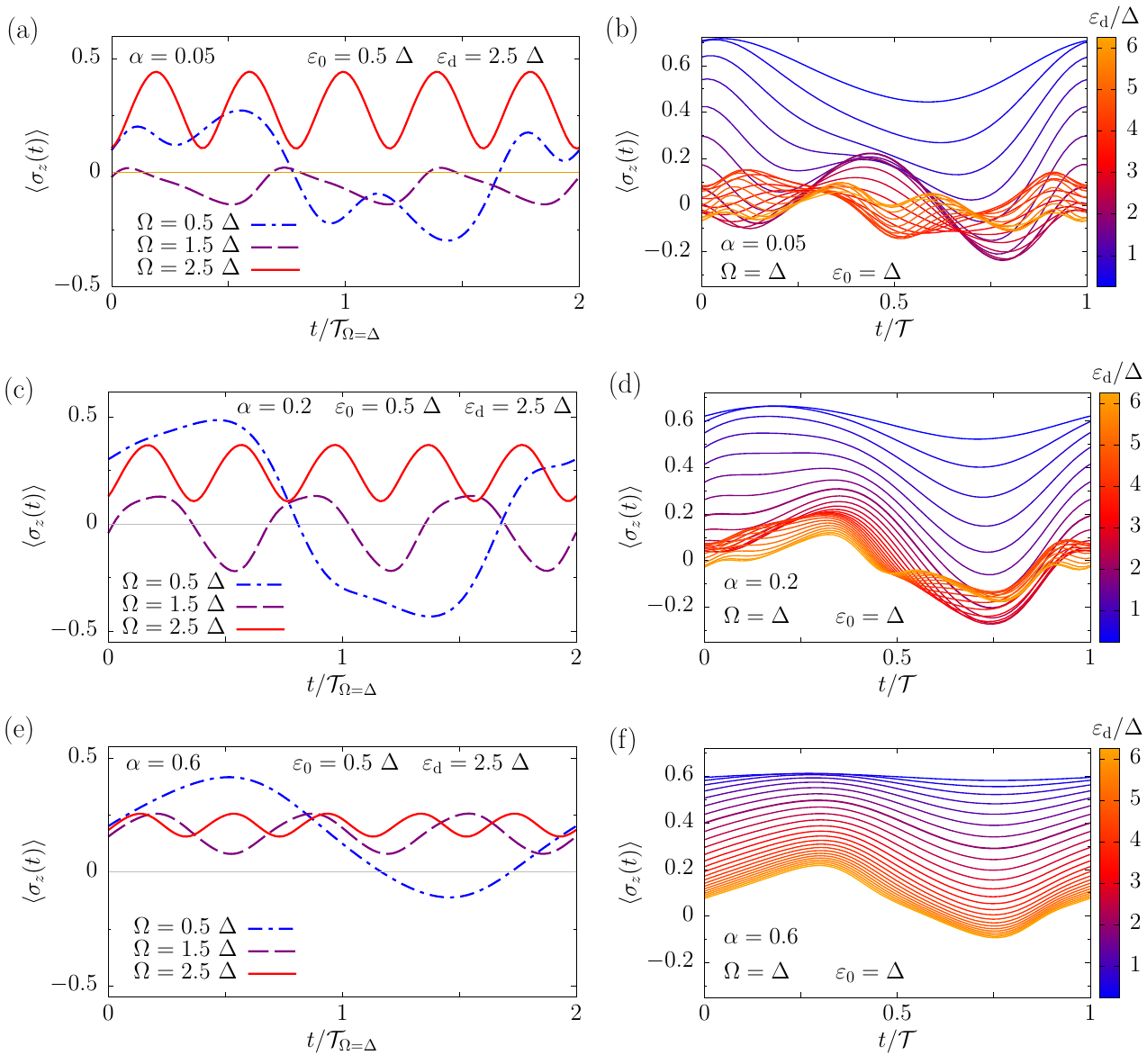}\caption{\footnotesize{Asymptotic Floquet states of the driven spin-boson model (11) 
with finite static bias. For each of the three values of the coupling strength $\alpha$, on the left panels the driving frequency $\Omega$ is varied at fixed driving 
strength $\varepsilon_{\rm d}=2.5~\Delta$ and for $\varepsilon_0=0.5~\Delta$, while the right panels show the asymptotic states at fixed frequency and constant 
static bias $\Omega=\varepsilon_0=\Delta$ for different driving amplitudes.
Temperature and Drude cutoff are $T=0.7~\hbar\Delta/k_B$ and $\omega_{\rm c}=5~\Delta$, respectively. Time is in units of the driving period $\mathcal{T}=2\pi/\Omega $ with $\Omega=\Delta$.}}
\label{biased}
\end{center}
\end{figure}

In the present section, we caclulate  asymptotic states of the driven spin-boson model for the three values of  coupling strength $\alpha=0.05,\;0.2$, and $0.6$ and different values of static bias $\varepsilon_0$, driving amplitude $\varepsilon_{\rm d}$, and frequency $\Omega$. All physical parameters are scaled with the 
frequency $\Delta$, the bare transition amplitude per unit time of the two-state system; see Eq.~(\ref{H}). 
The results presented are obtained for  temperature  $T=0.7~\hbar\Delta/k_B$ and Drude cutoff frequency $\omega_{\rm c}=5~\Delta$.
 For each value of $\alpha$, the coefficients of the embedding are determined by numerically optimizing  the approximated expressions for $h^{{\rm s/a}}(t)$,  Eq.~(\ref{happrox}), 
 against the numerical evaluations of the corresponding exact expressions in Eq.~(\ref{hsa}) (see Appendix~\ref{fit}).\\
\begin{figure}[ht!]
\begin{center}
\includegraphics[width=.95\linewidth,angle=0]{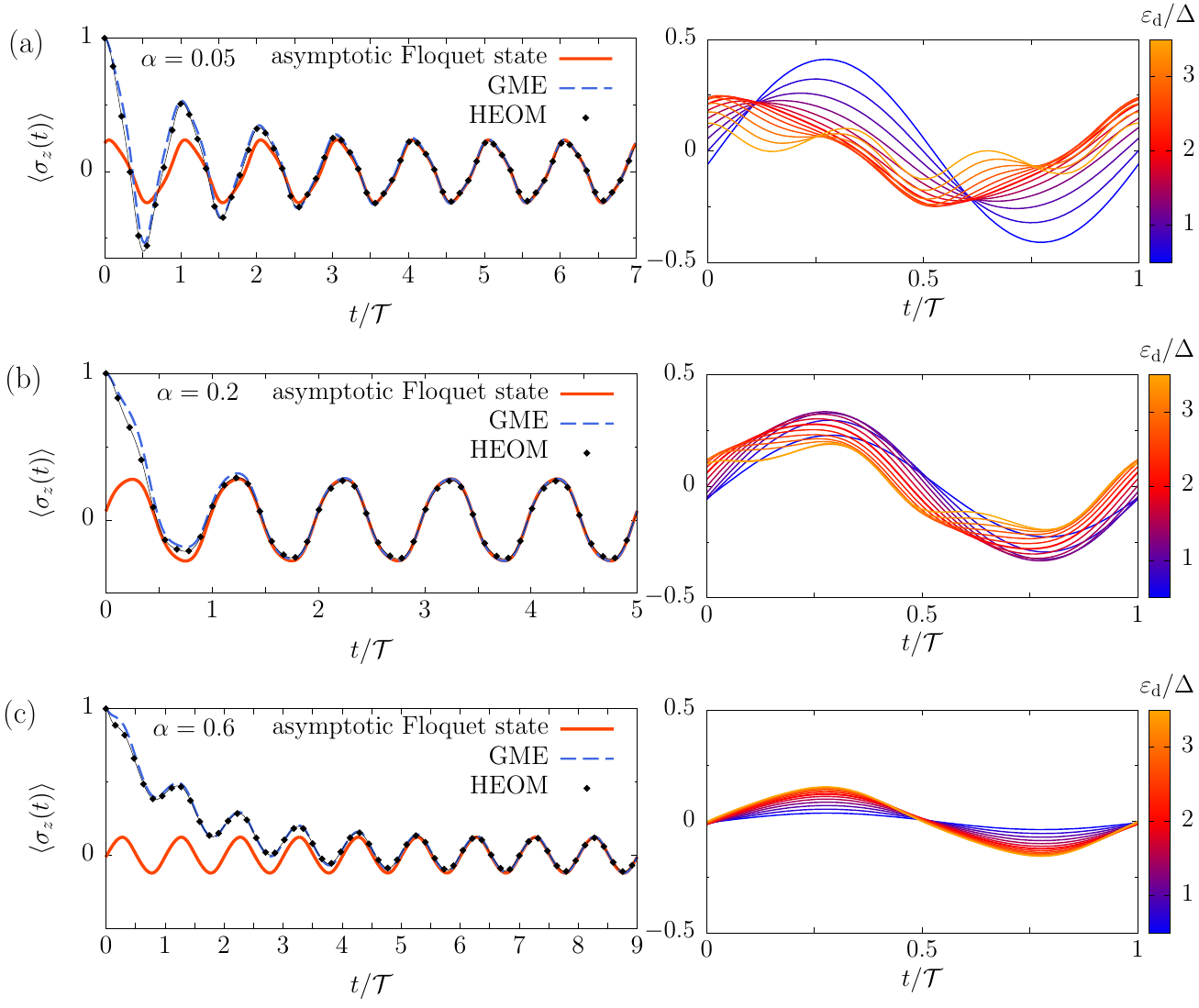}
\caption{\footnotesize{Asymptotic Floquet states of the driven spin-boson model, Eq. (11), 
for different coupling strengths and modulation  amplitudes. Left panels: Asymptotic population difference $\langle\sigma_z(t)\rangle=P_{+1}(t)-P_{-1}(t)$ {\it vs} 
time at fixed driving amplitude $\varepsilon_{\rm d}=2~\Delta$.  The populations $P_{\pm 1}(t)$ are the physical components of the extended system obeying the time local 
equation (\ref{system-compact}) (solid lines). 
A comparison is made with the corresponding propagated solutions of the GME~(\ref{GME}) (dashed lines) and the converged numerically exact 
HEOM (diamonds), starting with $P_{+1}(0)=1$. Right panels: Asymptotic Floquet states for different values of the drive amplitude $\varepsilon_{\rm d}$ 
evaluated for the same coupling strengths $\alpha$ as in the corresponding left panels. The thick red curves depict the same asymptotic dynamics as those in the corresponding left panels.
Note that, for different values of the spin-boson coupling strength $\alpha$, the maximum amplitude of the oscillations is reached at a different driving amplitude $\varepsilon_{\rm d}$. 
Fixed parameters are: Static bias $\varepsilon_0=0$, driving frequency $\Omega=\Delta$, temperature $T=0.7~\hbar\Delta/k_B$, and Drude cutoff $\omega_{\rm c}=5~\Delta$. 
Time is in units of the driving period $\mathcal{T}=2\pi/\Omega $.}}
\label{unbiased}
\end{center}
\end{figure}
\indent In Fig.~\ref{unbiased}, we depict the physical part of the asymptotic Floquet states of the extended system, namely the population 
difference $\langle\sigma_z(t)\rangle$ of the two-state system, at zero static bias, $\varepsilon_0=0$, and at fixed driving frequency $\Omega=\Delta$. This 
is done for the three values of coupling strength considered and by changing the drive amplitude. In the left panels we show, for $\varepsilon_{\rm d}=2~\Delta$, 
the asymptotic Floquet states over multiple driving periods along with the transient dynamics, as obtained by integrating the NIBA generalized master equation~(\ref{GME}) 
(see Appendix~\ref{HEOM} for details) with initial condition $\langle\sigma_z(0)\rangle=1$. This provides information on the time scales of relaxation to the asymptotic, 
time-periodic dynamics at different dissipation regimes and also certifies that the embedding procedure renders correctly the dynamics obtained from the GME. We note  
that, in the asymptotic limit, the results of the integration of the GME and of the Floquet solutions coincide perfectly, notwithstanding that the decomposition of the 
memory kernel employed in the embedding is only approximate, thus confirming the flexibility of the embeding method described in Sec.~\ref{embedding}. 
We benchmark the NIBA results with the numerically exact approach of  hierarchical equations of motion (HEOM) (see Appendix~\ref{HEOM}), obtaining reasonable 
agreement, even for $\varepsilon(t)\neq 0$, in all of the dissipation regimes.\\
\indent The right panels of Fig.~\ref{unbiased} depict the asymptotic Floquet states over a single driving period for driving strenghts $\varepsilon_{\rm d}$ in the range $[0.5~\Delta,3.5~\Delta]$. Note that the maximum amplitude of the oscillations is reached at different strengths for different values of $\alpha$ and that, in general, a large value of the coupling tends to suppress the higher harmonics which are most clearly visible for $\alpha=0.05$ in the nonlinear driving regime, i.e. when the condition $\varepsilon_{\rm d}/\Omega\ll 1$ is not met.\\  
\indent The same behavior is displayed by the system  also in the biased case, as depicted in the right panels of Fig.~\ref{biased}, where the Floquet asymptotic states over a time span of one diving period are shown with $\varepsilon_0=\Delta$, for driving strengths $\varepsilon_{\rm d}$ in the range $[0.25~\Delta,6.25~\Delta]$. We note that, as $\varepsilon_{\rm d}$ dominates over the static bias, the symmetry of the oscillations around $\langle \sigma_z\rangle=0$ tends to be restored. However, an increase of the coupling $\alpha$ tends to localize the spin state in the energetically more favourable state $|+1\rangle$, thus counteracting the effect of the time-periodic component of the driving.\\
\indent Figure~\ref{biased} also depicts the asymptotic states at finite static bias (left panels), $\varepsilon_0=0.5~\Delta$, for three values of the frequency $\Omega$. 
Note that, for $\alpha=0.05$, Fig.~\ref{biased}(a), even if the bias is positive, at the intermediate value of  frequency, $\Omega=1.5~\Delta$, the poupation difference 
mostly assumes negative values. This can be accounted for in terms of a negative effective bias $\varepsilon_{\rm eff}$, implicitly defined by the detailed balance relation~\cite{Grifoni1998,Magazzu2018} ${\rm K}^{\rm f}={\rm K}^{\rm b}\exp(\beta\hbar\varepsilon_{\rm eff})\;,$ where ${\rm K}^{\rm f (b)}=(1/\mathcal{T})\int_0^{\mathcal{T}}dt\;\int_0^{\infty}d\tau\;{\rm K}_{{+1-1 \above 0pt (-1+1)}}(t,t-\tau)$ are the  static rates obtained by averaging over the driving period the  time-integrated kernel matrix elements. 

\section{Conclusions and outlook}
In this work, we used the method developed in Ref.~\cite{Magazzu2017PRA} 
to find the asymptotic Floquet states of time-nonlocal evolution equations, 
to the periodically driven spin-boson model beyond the perturbative regime of coupling with the bosonic heath bath.\\ 
\indent The method rests on reshaping the memory kernel into a specific  form, which is suitable to model memory kernels which decay monotonously or with oscillations. This representation allows to embed the system dynamics into an enlarged state space, by coupling the actual physical variables to a set of auxiliary non-physical variables. 
The dynamics of the so-obtained enlarged system is governed by a time-local equation to which the standard Floquet formalism applies.\\
\indent We considered temperature/coupling regimes where the non-interacting blip approximation of the path integral 
expression for the reduced dynamics  describes satisfactorily the driven spin-boson model, 
as we demonstrate upon comparison with evaluations of the dynamics by using converged hierarchical 
equations of motions. In particular, the probabilities associated to the spin states are given by a  generalized master 
equation displaying a memory kernel which accounts for the effects of the environment and of the driving. 
By using an optimization procedure, we performed  an approximate embedding and find the asymptotic Floquet states 
of the spin dynamics as the projection of the Floquet states of the enlarged system onto the physical subspace. 
We illustrated this idea by calculating asymptotic Floquet states in different dissipation and driving regimes, both for the unbiased and biased system.\\
\indent Our results demonstrate, see Fig~\ref{unbiased}, that the proposed method still performs well in the cases where the expression in Eq.~(\ref{MK}) is employed with a finite number of terms to approximate the memory kernel. 
Finally, the study of the driven dynamics of the spin-boson model performed in the present work can be generalized to multi-site or multi-level systems subjected to periodical modulations~\cite{Grifoni1996, Grifoni1998, Thorwart2001,Magazzu2015, MagazzuJSTAT2016}.

\section*{Acknowledgements}
S. D. and P. H. acknowledge the support by the Russian Science Foundation, Grant No.~15-12-20029 (S.D.)
and  by the Deutsche Forschungsgemeinschaft (DFG) via the grants
DE1889/1-1 (S.D.) and HA1517/35-1 (P.H.).

\appendix

\section{HEOM and details of the simulations}\label{HEOM}
\begin{figure}[t]
\begin{center}
\includegraphics[width=0.45\linewidth,angle=0]{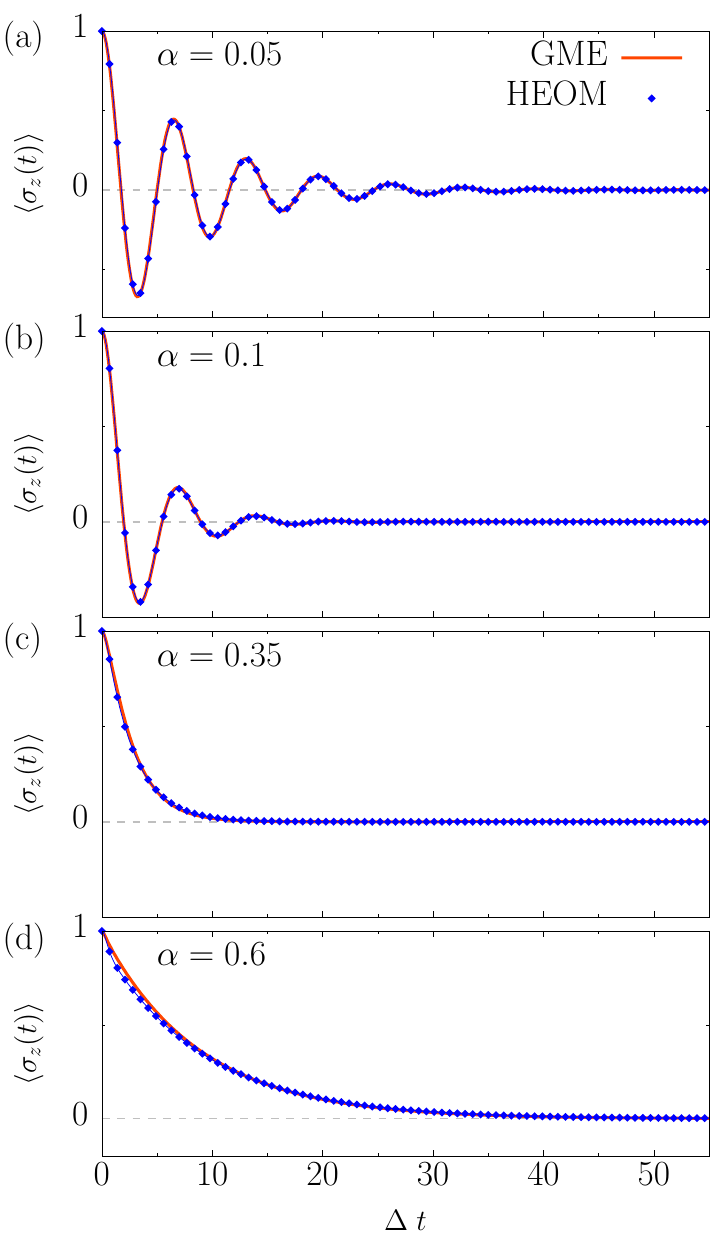}\\
\caption{\footnotesize{Comparison between the results of the GME~(\ref{GME}) (solid lines) and converged HEOM (diamonds) 
in the unbiased, static case $\varepsilon(t)=0$, for different values of the coupling strength $\alpha$. Temperature and cutoff frequency are  $T=0.7~\hbar\Delta/k_B$ and $\omega_{\rm c}=5~\Delta$, respectively.}}
\label{comparison}
\end{center}
\end{figure}

Consider an open system of effective mass $M$ coupled to a heat bath with Ohmic-Drude spectral density function and cutoff frequency $\omega_{\rm c}$,
\begin{eqnarray}\label{}
J(\omega)=\eta\omega(1+\omega^2/\omega_{\rm c}^2)^{-1}\;,
\end{eqnarray}
where $\eta=M\gamma$ is the viscosity parameter and  $\gamma$ the friction parameter with dimensions of frequency. 
Specialization to a two-state system characterized by tunneling between sites separated by the distance  $q_0=2d$ yields~\cite{Weiss2012}: 
\begin{eqnarray}\label{}
\alpha=\frac{\eta q_0^2}{2\pi\hbar}=\frac{2}{\pi}\eta\frac{d^2}{\hbar}\;. \end{eqnarray}
We set $d=1$, in units of $(M\Delta/\hbar)^{-1/2}$, so that the position eigenvalues are identified with the eigenvalues of $\sigma_z$.\\
\indent In the HEOM approach~\citep{Tanimura1989,Ishizaki2005,Tanimura2006}, the time evolution of an open system interacting 
through the operator $V$ (in the present case $V\equiv\sigma_z$)  with a harmonic heat bath 
is governed  by a set of coupled time-local evolution equations for a set of  density matrices $\{\rho_{n,\;\mathbf{j}}(t)\}$ 
of which only $\rho_{0,\;\mathbf{0}}(t)$ describes the actual physical system, 
while the others are auxiliary, non-physical density matrices. This approach exploits the fact that, for a Ohmic spectral density with Drude cutoff, the bath correlation function 
\begin{eqnarray}\label{L}
L(t)=\frac{1}{\pi}\int_0^\infty d\omega\; J(\omega)\left[\coth\left(\frac{\beta\hbar\omega}{2}\right)\cos(\omega t)-{\rm i}\sin(\omega t)\right]
\end{eqnarray}
can be expressed as a series over the 
Matsubara frequencies $\nu_k=2\pi k/(\hbar\beta)$, as done in Eq.~(\ref{ReQ}). The Pad\'e spectrum decomposition allows for an efficient truncation of this series to the first ${\rm M}$ terms~\cite{Tsuchimoto2015}, even when the temperature is not much larger than the characteristic frequency $\Delta$ of the system, as in our case where $k_BT<\hbar \Delta$. The hierarchy of equations reads
\begin{eqnarray}\label{hierarchy}
\dot\rho_{n,\;\mathbf{j}}(t)&=&-\left[\frac{{\rm i}}{\hbar} H(t)^{\times} +n \omega_{\rm c}  +\sum_{k=1}^{\rm M}  j_k\nu_k + \left( g-\sum_{k=1}^{\rm M}f_k\right) V^\times V^\times \right] \rho_{n,\;\mathbf{j}}(t)\nonumber\\
&&-{\rm i}V^\times \rho_{n+1,\;\mathbf{j}}(t)-n\omega_{\rm c} (\theta V^\circ+{\rm i}\varphi V^\times)\rho_{n-1,\;\mathbf{j}}(t)\nonumber\\
&&-{\rm i}\sum_{k=1}^{\rm M} V^\times \rho_{n,\;\mathbf{j}_{k+}}(t) -{\rm i}\sum_{k=1}^{\rm M}j_k\nu_kf_k V^\times  \rho_{n,\;\mathbf{j}_{k-}}(t)\;,
\end{eqnarray}
where $\mathcal{O}^{\times}\rho :=[\mathcal{O},\rho]$ and $\mathcal{O}^{\circ}\rho :=\{\mathcal{O},\rho\}$, for a given operator $\mathcal{O}$, and where the following quantities have been introduced
\begin{eqnarray}
\theta=\frac{\eta\omega_{\rm c}}{2}\;,\quad \varphi=\theta\cot\left(\frac{\beta\hbar\omega_{\rm c}}{2}\right),\quad g=\frac{\eta}{\hbar\beta}-\varphi,\quad{\rm and}\quad f_k=\frac{\eta}{\hbar\beta}\frac{2\omega_{\rm c}^2}{\nu_k^2-\omega_{\rm c}^2}\;.
\end{eqnarray}
 It is understood that the contributing density matrices have both $n$ and all of the components of the vector index $\mathbf{j}=(j_1,\dots, j_k,\dots,j_{\rm M})$ nonnegative. Moreover $\mathbf{j}_{k\pm}$ is a short-hand notation for $(j_1,\dots, j_k\pm 1,\dots,j_{\rm M})$.
The terminators of the hierarchy of equations~(\ref{hierarchy}) are identified by the condition $n+\sum_{k=1}^{\rm M} j_k={\rm N}$, for some ${\rm N}\gg \omega_0/{\rm min}(\omega_{\rm c},\nu_1)$, and read
\begin{eqnarray}\label{terminators}
\dot\rho_{n,\;\mathbf{j}}(t)\simeq -\left[ \frac{{\rm i}}{\hbar} H(t)^{\times} + \left( g-\sum_{k=1}^{\rm M}f_k 	\right) V^\times V^\times \right] \rho_{n,\;\mathbf{j}}(t)
\end{eqnarray}
\indent In Figs.~\ref{unbiased} and~\ref{comparison}, NIBA results for the  reduced dynamics are compared with those from converged HEOM for the driven and non-driven case, respectively.
Simulations performed by implementing the HEOM with Pad\'e spectrum decomposition of the bath correlation function are converged with ${\rm M}=3$ and ${\rm N}=10$, for all the parameters considered. 
The propagation of the HEOM is obtained by using a second order Runge-Kutta scheme with time step $\delta t=5\times 10^{-4}~\Delta^{-1}$. 
The GME~(\ref{GME}) is propagated by using the trapezoid rule for the integral and the forward finite difference 
for the time derivative, according to the simple implicit scheme described in~\cite{MagazzuJSTAT2016} for 
a multi-level generalization of the spin-boson model. The time step is  $\delta t=5\times 10^{-3}~\Delta^{-1}$.
 
\section{Fits to kernels}\label{fit}
\begin{figure}[t]
\begin{center}
\includegraphics[width=0.45\linewidth,angle=0]{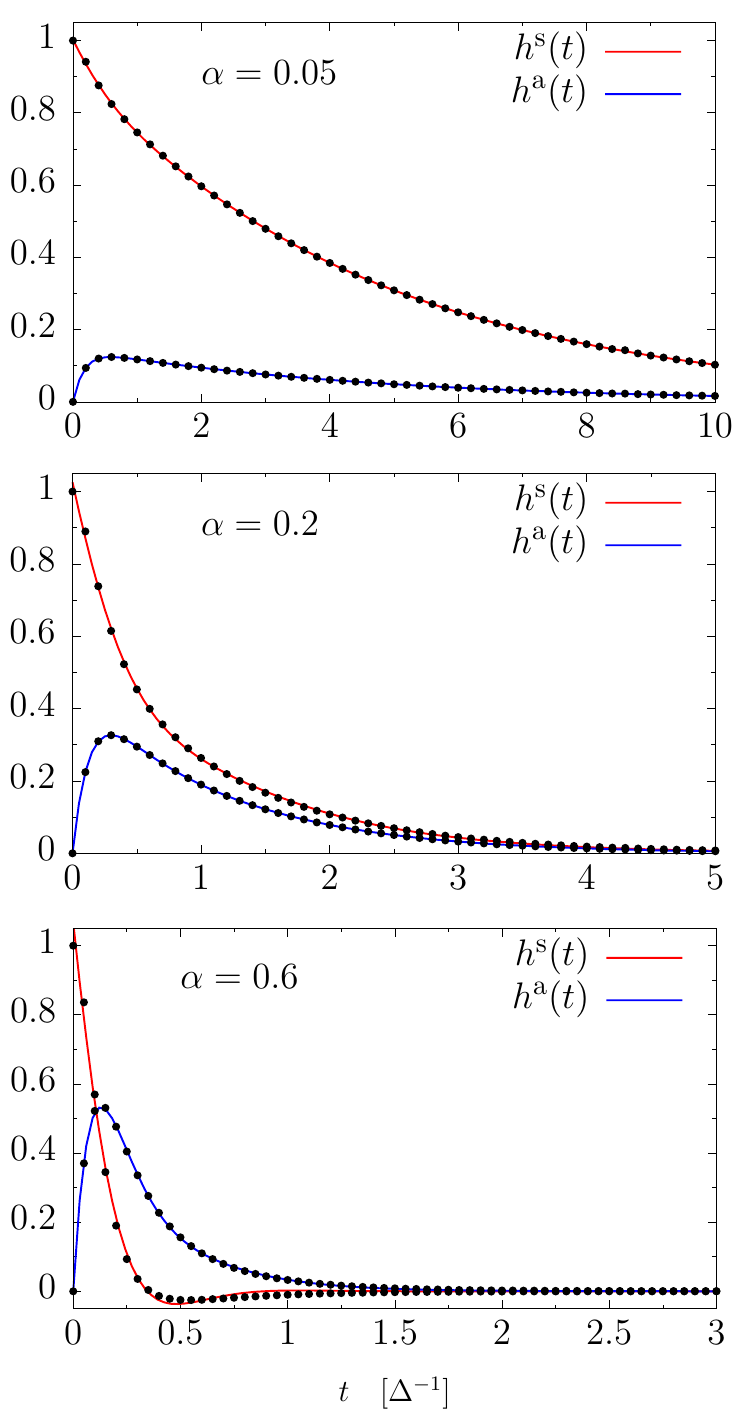} \caption{\footnotesize{Numerical fits of the functions $h^{\rm s/a}(t)$ defined in Eq.~(\ref{hsa}) 
performed by using  the approximated expressions in Eq.~(\ref{happrox})}. The coefficients of these approximated expressions (solid lines) 
are determined by fitting the numerically exact evaluations (bullets). The latter are performed by truncating the sum for $Q'(t)$ to the first $4000$ 
terms (see Eq.~(\ref{Q2})). Temperature and Drude cutoff frequency are $T=0.7~\hbar\Delta/k_B$ and $\omega_{\rm c}=5~\Delta$, respectively.}
\label{fig-fit}
\end{center}
\end{figure}

In Fig.~\ref{fig-fit} we show the optimized curves corresponding 
to the approximate expressions in Eq.~(\ref{happrox}) with coefficients 
extracted by fitting the numerically exact evaluations of the functions in  Eq.~(\ref{hsa}). 
Note that, as these functions do not depend on the driving, the results shown in the present work required three sets of coefficients, corresponding to the three panels in Fig.~\ref{fig-fit}.

\end{document}